**Title:**

Total-Body Parametric Imaging Using Relative Patlak Plot

**Authors:**

Siqi Li[1], Yasser G. Abdelhafez[1], Lorenzo Nardo[1], Simon R. Cherry[2,1], Ramsey D. Badawi[1,2], Guobao Wang[1]

**Affiliations:**

1. Department of Radiology, University of California – Davis
2. Department of Biomedical Engineering, University of California – Davis

**First author and corresponding author**:

Siqi Li, Department of Radiology, University of California, Davis. 4860 Y Street, Sacramento, CA 95817. Email: sqlli@ucdavis.edu. Phone: 530-953-1662.

**Word count**: 5891

**Figures**: 7 (supplemental figures: 2)

**Tables**:1 (supplemental table: 1)

# ABSTRACT

The standard Patlak plot, a simple yet efficient model, is widely used to describe irreversible tracer kinetics for dynamic PET imaging. Its widespread application to whole-body parametric imaging remains constrained due to the need for a full-time-course input function (e.g., 1 hour). In this paper, we demonstrate the *Relative Patlak (RP) plot*, which eliminates the need for the early-time input function, for total-body parametric imaging and its application to 20-minute clinical scans acquired in list-mode. **Method:** We conducted a theoretical analysis to indicate that the RP intercept $b'$ is equivalent to a ratio of the standardized uptake value (SUV) relative to the plasma concentration, while the RP slope $K_i'$ is equal to the standard Patlak $K_i$ (net influx rate) multiplied by a global scaling factor for each subject. One challenge in applying RP to a short scan duration (e.g., 20 minutes) is the resulting high noise in the parametric images. We applied a self-supervised deep kernel method for noise reduction. Using the standard Patlak plot as the reference, the RP method was evaluated for lesion quantification, lesion-to-background contrast, and myocardial visualization in total-body parametric imaging in 22 human subjects (12 healthy subjects and 10 cancer patients) who underwent a 1-hour dynamic $^{18}$F-FDG scan. The RP method was also applied to the dynamic data reconstructed from a clinical standard 20-minute list-mode scan either at 1-h or 2-h post-injection for two cancer patients. **Result**: We demonstrated that it is feasible to obtain high-quality parametric images from 20-minute scans using RP parametric imaging with a self-supervised deep-kernel noise reduction strategy. The RP slope $K_i'$ was highly correlated with the standard Patlak $K_i$ in lesions and major organs, demonstrating its quantitative potential across subjects. Compared to conventional SUVs, the $K_i'$ images significantly improved lesion contrast and enabled visualization of the myocardium for potential cardiac assessment. The application of the RP parametric imaging to the two clinical scans also showed similar benefits. **Conclusion**: Using total-body PET with the RP approach, it is feasible to generate parametric images using data from a 20-minute clinical list-mode scan.

## INTRODUCTION

Dynamic $^{18}$F-fluorodeoxyglucose (FDG) positron emission tomography (PET) with tracer kinetic modeling enables multiparametric imaging and provides more accurate metabolic information as compared to static imaging (*1–4*). Among various kinetic modeling approaches, the Patlak graphical plot (*5*) is a commonly used linear kinetic model to describe FDG kinetics. The slope parameter of the standard Patlak plot, $K_i$, represents FDG net influx rate and has demonstrated advantages beyond the standardized uptake value (SUV) images, for example, for improving tumor detection and discrimination (*6–9*), monitoring response to treatment (*10–13*) and evaluating the prognostic outcome (*14–16*).

Whole-body parametric imaging with the standard Patlak plot has been implemented on conventional PET scanners with a short axial field-of-view (AFOV) ranging from 15-30 cm using a multi-bed and multi-pass acquisition strategy (*9,17*). The advent of long AFOV and total-body PET scanners with a much longer AFOV (>1 meter), such as the UIH uEXPLORER (*18–20*) and the Siemens Quadra (*21*) has further simplified and improved the implementation of Patlak parametric imaging because of much-improved detection sensitivity and simultaneous coverage of multiple organs. However, all these methods require a full-time-course dynamic scan (e.g., 1 hour) to obtain an image-derived input function, which limits their broad use. Although a population-based input function may be used (*22–24*), it is challenging to adapt the approach to individual patients, particularly for pediatric patients.

The *Relative Patlak (RP) plot* (*25*), which does not require the early-time input function but only the late-time input function, is another solution to streamline the parametric imaging process. This method has been recently deployed on commercial short-AFOV scanners for whole-body parametric imaging (*17*). The RP slope $K_i'$ is equivalent to the standard Patlak $K_i$ multiplied by a global scaling factor, thus providing comparable performance for lesion detection (*25*). However, its potential for absolute quantitation has not been demonstrated. Early implementation of the RP plot also commonly used a scan duration of 30 minutes or more (*17*), requiring data acquisition longer than a typical clinical scan duration (commonly up to 20 minutes) (*26*).

This paper aims to develop and evaluate the potential of the RP plot for total-body parametric imaging from the dynamic data of a clinical 20-minute scan. This method will add a new ability on top of standard clinical imaging to generate parametric images from the same acquired list-mode data. One challenge here is that the noise level in the resulting parametric images will be higher due to the use of a shorter scan duration. We thus propose a self-supervised deep kernel method to improve the quality of RP parametric imaging. The proposed total-body RP approach has the unique advantage of being usable for delayed scans (e.g., a 20-minute scan at 1-2 h post-injection) and for pediatric patients, for whom it is difficult to obtain a population-based input function.

## METHODS AND MATERIALS

### Standard and Relative Patlak Plots

The standard Patlak plot (*5*) describes the linear relationship between the normalized tissue concentration and normalized integral of input function after an equilibrium time $t^*$ with the following equation:

$$\frac{C_{\mathrm{T}}(t)}{C_{\mathrm{p}}(t)} = K_{\mathrm{i}} \cdot \frac{\int_0^t C_{\mathrm{p}}(\tau)d\tau}{C_{\mathrm{p}}(t)} + b \quad (t > t^*), \tag{Eq. 1}$$

where $C_{\mathrm{T}}(t)$ denotes the FDG concentration in a tissue region at the time $t$. $C_{\mathrm{p}}(t)$, also called input function, represents the FDG concentration in the plasma at the time $t$. $K_{\mathrm{i}}$ is a slope parameter representing the FDG net influx rate. $b$ is an intercept parameter that is typically considered as a mixture of blood volume and free-state FDG volume of distribution in tissue (*5*). The two parameters can be estimated via linear regression. The common choice of $t^*$ for total-body Patlak parametric imaging is greater than or equal to 30 minutes (*27,28*). Note that although only the time activity data of later than $t^*$ is required for $C_{\mathrm{T}}(t)$, the plot still requires a full-time-course input function from the injection time to the end of the scan for $C_{\mathrm{p}}(t)$.

The Relative Patlak (RP) plot (*25*) was proposed as formulated in the following equation:

$$\frac{C_{\mathrm{T}}(t)}{C_{\mathrm{p}}(t)} = K_{\mathrm{i}}' \cdot \frac{\int_{t^*}^t C_{\mathrm{p}}(\tau)d\tau}{C_{\mathrm{p}}(t)} + b' \quad (t > t^*), \tag{Eq. 2}$$

where $K_{\mathrm{i}}'$ and $b'$ are the new RP slope and intercept. Compared to the standard Patlak plot, the RP plot eliminates the need for the early-time input function from 0 to $t^*$ and only requires the late-time input function from $t^*$ to the end of the scan, thus potentially providing a more efficient approach for parametric imaging.

**Theoretical Interpretation of the Relative Patlak Plot**

Our earlier study (*25*) demonstrated that the RP slope $K_{\mathrm{i}}'$ is equivalent to the standard Patlak $K_{\mathrm{i}}$ multiplied by a global scaling factor $\alpha$ in each subject:

$$K_{\mathrm{i}}' = \alpha K_{\mathrm{i}}, \tag{Eq. 3}$$

leading to an equivalent spatial distribution between the parametric images of the two slopes.

The interpretation of the RP intercept $b'$, however, remains under-explored in previous studies. We note that Eq. 2 leads to the following equation by setting $t = t^*$:

$$\frac{C_{\mathrm{T}}(t^*)}{C_{\mathrm{p}}(t^*)} = K_{\mathrm{i}}' \cdot \frac{\boxed{\int_{t^*}^{t^*} C_{\mathrm{p}}(\tau)d\tau = 0}}{C_{\mathrm{p}}(t)} + b' = b', \tag{Eq. 4}$$

which indicates that the intercept $b'$ is equivalent to the SUV ratio (SUVr) relative to the plasma input function at the time $t^*$:

$$b' = \text{SUVr}(t^*) \equiv \frac{C_\text{T}(t^*)}{C_\text{p}(t^*)}. \tag{Eq. 5}$$

This equivalence suggests the RP intercept $b'$ is not arbitrary but has a physiological interpretation, further expanding the theoretical aspects of the RP plot.

**RP Parametric Imaging with Self-Supervised Deep-Kernel Denoising**

One potential application of the RP plot is to shorten the scan duration for parametric imaging by increasing $t^*$. Our preliminary analysis shown in Supplemental Figure. 1 suggests that the noise level of $K'_i$ image will be higher if the scan duration is shortened. We thus applied the deep-kernel method (*29*) to overcome the noise issue.

A denoised image $x_{denoised}$ can be modeled using a generalized kernel representation:

$$x_{denoised} = \mathcal{K}(\boldsymbol{\theta}; \boldsymbol{Z})x_{noise}, \tag{Eq. 6}$$

which may also be explained as a type of non-local mean denoising (*30*). $x_{noise}$ is a noisy image. $\mathcal{K}(\boldsymbol{\theta}; \boldsymbol{Z})$ is a kernel matrix built on the image prior data $\boldsymbol{Z}$ with $\boldsymbol{\theta}$ including any parameters that determine the kernel representation. Unlike the conventional kernel method (*31*) that uses an empirically defined $\boldsymbol{\theta}_e$, the deep kernel method (*29*) enables an optimized and learned kernel matrix by extending the kernel representation (Eq. 6) into trainable neural networks. The kernel matrix $\mathcal{K}_l(\boldsymbol{\theta}_l; \boldsymbol{Z})$ now includes the trainable parameter $\boldsymbol{\theta}_l$ and will be trained using a simplified self-supervised strategy for deep kernel learning through a denoising auto-encoder framework:

$$\widehat{\boldsymbol{\theta}}_l = \text{argmin}_{\boldsymbol{\theta}_l} \sum_{m=1}^{n_z} ||z_a - \mathcal{K}_l(\boldsymbol{\theta}_l; \boldsymbol{Z})z_m||^2, \tag{Eq. 7}$$

where $\boldsymbol{Z} = \{z_m\}_{m=1}^{n_z}$ consists of $n_z$ dynamic frames. $z_a$ is the corresponding mean image of all the frames and is considered as the clean training label. In this work, $n_z = 4$ and each $z_m$ was thus a 5-minute frame. All the training settings, including neighborhood number of 200, training iteration of 300, the learning rate of 1e-3, and initialization were the same as used in (*29*). Once $\widehat{\boldsymbol{\theta}}_l$ is trained, $\mathcal{K}_l(\widehat{\boldsymbol{\theta}}_l; \boldsymbol{Z})$ is then used on both dynamic activity images and parametric images for noise reduction.

**Total-body Dynamic PET Data Acquisition for Validation**

Twenty-two subjects, including twelve healthy volunteers and ten patients with immunotherapy naïve, metastatic genitourinary cancer (GUC), were included in this study and scanned using the uEXPLORER total-body PET scanner. Prior Ethics Committee/IRB approval and informed consent were obtained for these subjects. After a total-body CT scan, each participant underwent a 1-hour dynamic scan with an injection of approximately 370 MBq $^{18}$F-FDG. The resulting list-mode data was reconstructed into 29 frames of dynamic images (6×10s,

2×30s, 6×60s, 5×120s, 4×180s, 6×300s) using vendor-implemented time-of-flight ordered-subset expectation maximization (OSEM) algorithm with 4 iterations and 20 subsets (*20*). All standard corrections, including random correction, scatter correction, attenuation correction, deadtime correction and decay correction, were applied. The image size of each frame was 150×150×486 and the voxel size was 4×4×4 mm$^3$.

The RP plot was implemented on the late 20-minute data of each 1-hour dynamic scan (i.e., $t^* = 40$-minute) with an image-derived input function from a region-of-interest (ROI) placed in the ascending aorta region. The standard Patlak plot was also applied to the same 20-minute dynamic data but with a full 1-hour blood input function. The SUV image was calculated using the last 5 minutes of data. A total of 26 lesions were identified in GUC patients; ROIs encompassing these lesions were drawn using AMIDE software (*32*). In addition, an ROI was also placed in major organs, including the liver, lung, spleen, spine bone marrow (an average of ten spine sections across thoracic, lumbar, and sacrum vertebrae), pelvic bone marrow (an average of four ROIs, two on the left and two on the right), thigh muscle (an average of two ROIs of left and right thighs), and grey matter, leading to 154 organ ROIs in all subjects.

**Demonstration of RP Parametric Imaging with Deep-Kernel Noise Reduction**

The RP parametric imaging was implemented using different reconstruction/denoising methods, including the standard OSEM method, conventional kernel method (*31*), 4D deep image prior (DIP) method (*33*) and the deep-kernel denoising method. Note that the learning-based methods here are all based on single subjects and do not require population-based pretraining. Different approaches were compared for parametric imaging of the RP $K_i'$ using the trade-off of lesion contrast recovery ($K_i'$ absolute value) versus the background noise level calculated as the standard deviation of voxels in the liver ROI.

With the deep-kernel noise reduction approach, we then verified the total-body global-scaling relationship between the RP slope $K_i'$ image and the standard Patlak slope $K_i$ image, as well as the equivalence between the RP intercept $b'$ and the SUVr at $t^*$ using the scatter plots of all image voxels between the two methods. Linear regression was used to evaluate the correlation.

**Comparison of Total-Body $K_i'$ Images with $K_i$ and SUV for Lesion Contrast and Myocardial Visualization in Cancer Patients**

To demonstrate the potential benefit of total-body RP parametric imaging for lesion detection, the parametric images of $K_i'$ were compared to the SUV images using the standard Patlak slope $K_i$ images as the reference. We computed the lesion-to-liver contrast ratio (CR) for $K_i'$, SUV and $K_i$ for all 26 lesions from 10 cancer patients. The paired t-test and Wilcoxon signed-rank test were used to evaluate their statistical differences. *P* values of less than 0.05 were considered statistically significant. The CR difference between $K_i'$ and SUV, defined as $(\mathrm{CR}_{K_i'} - \mathrm{CR}_{\mathrm{SUV}})/\mathrm{CR}_{\mathrm{SUV}} \times 100\%$, was calculated for each lesion. A positive value means the better contrast provided by $K_i'$. In addition, parametric imaging has the advantage of better visualizing the myocardium than the SUV does, as shown in our early work (*34*). Thus, the three images were also compared for their ability to visualize the myocardium in cancer patients.

**Evaluation of the Quantitative Potential of $K_i'$ using $K_i$ as the Reference**

Because there is a global scaling factor between the parametric image of RP $K_i'$ and that of the standard Patlak $K_i$ in each subject, it was unclear whether the RP $K_i'$ can be quantitative across subjects. We conducted a linear correlation analysis between $K_i'$ and $K_i$ using the data of 154 organ ROIs and 26 lesion ROIs from the 22 subjects. The coefficient of variation, defined as the ratio of the standard deviation to the mean, was used to measure how discretely this global scaling factor $\alpha$ is between different subjects. In addition, we also estimated the global scaling factor from 12 healthy subjects and then applied it to the $K_i'$ images of 10 cancer patients for calibration. The difference in all 26 lesion ROIs between calibrated $K_i$ and reference $K_i$ was quantified using the Bland-Altman plot.

**Application to Clinical List-mode Scans**

We further applied the RP parametric imaging approach to two clinical 20-minute list-mode FDG-PET scans to demonstrate the potential of the method. One was for a lymphoma patient scanned from 60-80 minutes post-injection, and the other was for a lung cancer patient scanned from 120-140 minutes post-injection. The list-mode raw data of each scan was reprocessed to generate four 5-minute frames of dynamic data. Implementation details and reconstruction settings were the same as those used in the aforementioned validation study. Because there is no early-time input function data available, the standard Patlak plot was not applicable. Thirteen lesions (eight from the lymphoma patient and five from the lung cancer patient) were identified and used to evaluate the lesion contrast for $K_i'$ and SUV. The paired t-test was used to indicate statistical significance. All human subjects' basic characteristics are summarized in Supplemental Table 1.

**RESULTS**

**Total-Body RP Parametric Imaging using Deep-Kernel Noise-Reduction**

Figure 1 shows the total-body parametric image of the RP slope $K_i'$ (mL/min/cm$^3$) generated from a 20-minute dynamic scan for (A) a healthy subject and (B) a cancer patient using the standard OSEM reconstruction method and the deep-kernel noise reduction method. Without post-reconstruction noise reduction, the standard OSEM yielded a noisy $K_i'$ image. The deep kernel method substantially improved the $K_i'$ image quality with lower noise and clearer lesion visualization,

Comparison of the deep-kernel method with the conventional kernel method, and 4D DIP method noise-reduction methods are presented in Supplemental Figure 2. While the latter two methods both suppressed the noise, they also significantly reduced lesion contrast. The deep-kernel method achieved a similar high lesion contrast as the OSEM did but also a much lower background noise as the 4D deep image prior had.

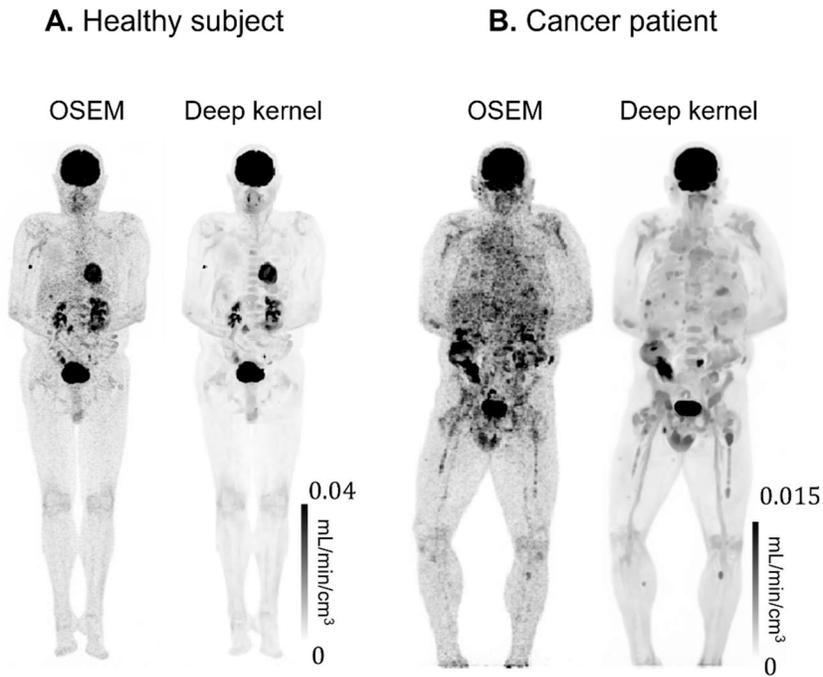

**FIGURE 1** Total-body RP $K_i'$ images from a 20-minute dynamic FDG scan (40-60 minutes post-injection) based on the standard OSEM reconstruction and the deep-kernel noise reduction method for (A) a healthy subject and (B) a cancer patient. Images are shown in maximum intensity projection.

**Demonstration of the Theoretical Aspects of Total-Body RP Parametric Imaging**

Figure 2A shows the comparison between the standard Patlak slope $K_i$ and RP slope $K_i'$ images for a cancer patient. These two images were visually identical, though their absolute values are different. The correlation plot of all the image voxels in Figure 2B verified that $K_i'$ is equivalent to $K_i$ multiplied by a global scaling factor, i.e., $\alpha$=1.3 for this subject. The y-intercept of the fitted line was nearly zero.

Figure 3A shows the comparison between the SUVr ($t^*$=40-minute) and RP intercept $b'$ images for the same subject. The $b'$ image was closely equivalent to the SUVr image with an excellent fitting by the identity line. The linear correlation was statistically significant with a high correlation coefficient R (close to 1) and a minimal P value (<0.0001).

Tests on other subjects also showed similar results. These data together verified the theoretical relationships of the RP slope $K_i'$ and intercept $b'$ with respect to the standard Patlak slope and SUVr, respectively.

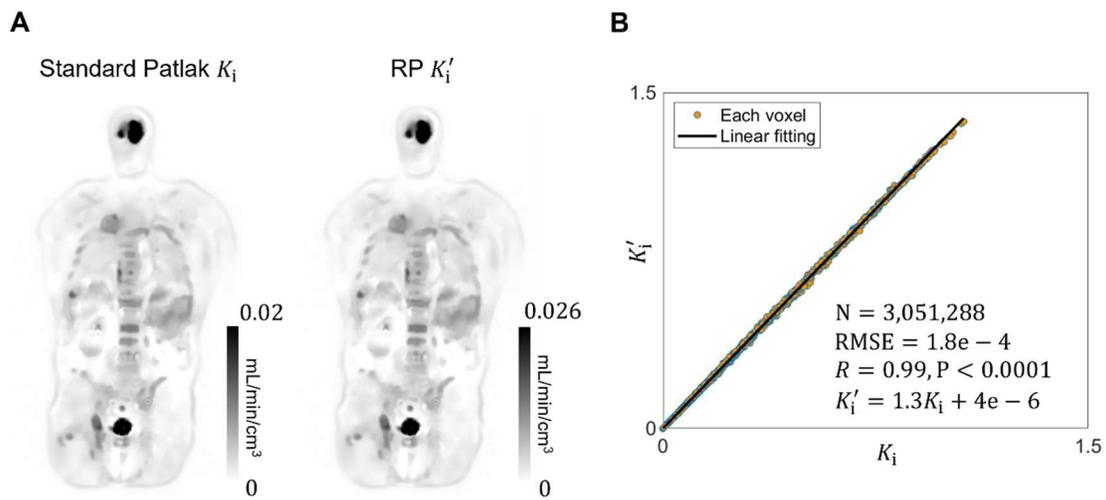

**FIGURE 2**. Demonstration of the relationship between standard Patlak $K_i$ and RP $K_i'$ for a cancer patient. (A) Total-body $K_i$ and $K_i'$ images. (B) Correlation plot of all voxels in the images. The global scaling factor was approximately 1.3. The correlation coefficient (R) was close to 1.

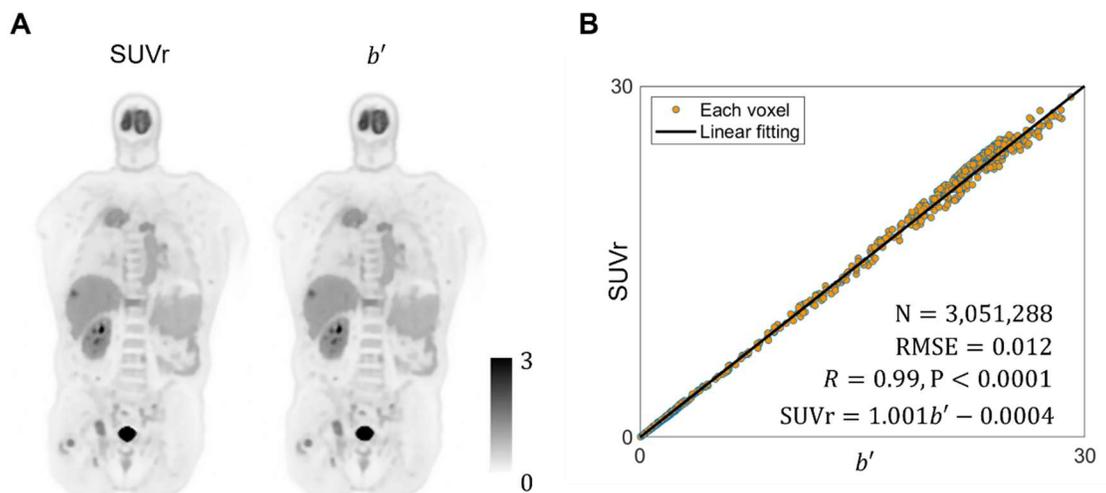

**FIGURE 3**. Demonstration of the equivalence between SUVr ($t^*$ = 40-minute) and RP intercept $b'$ in a cancer patient. (A) Total-body images of SUVr and $b'$; (B) Correlation plot of all image voxels. The fitting line was almost identical to the identity line. The correlation coefficient (R) was close to 1.

### Comparison of RP $K_i'$ with SUV for Lesion Contrast and Myocardial Visualization

Figure 4A shows the image comparison between SUV, $K_i$ and $K_i'$ for a cancer patient. All images were superimposed on CT. A follow-up contrast CT image was included to confirm a lung metastasis indicated by the arrow. The lesions were clearly identified in both $K_i$ and $K_i'$ images, while the signal was much weaker in the SUV image. The $K_i'$ image showed a lesion-to-liver contrast of 2.5, which is the same as $K_i$ provided but much higher than the 0.59 provided by the SUV image. Figure 4B further shows a group comparison of the lesion-to-liver contrast of all 26 lesions in the ten cancer patients. $K_i'$ had a nearly 4-fold higher lesion contrast than SUV, with a median value of 4.85 versus 1.22. As indicated by the horizontal paired lines, $K_i$ and $K_i'$ had the exactly same contrast results because the global scaling effect does not change the contrast. The contrast ratio difference between $K_i'$ and SUV for each individual lesion is further shown in Figure 4C. In 25 out of the 26 lesions, the $K_i'$ demonstrated a higher contrast, while the remaining one showed a slightly lower contrast than the SUV.

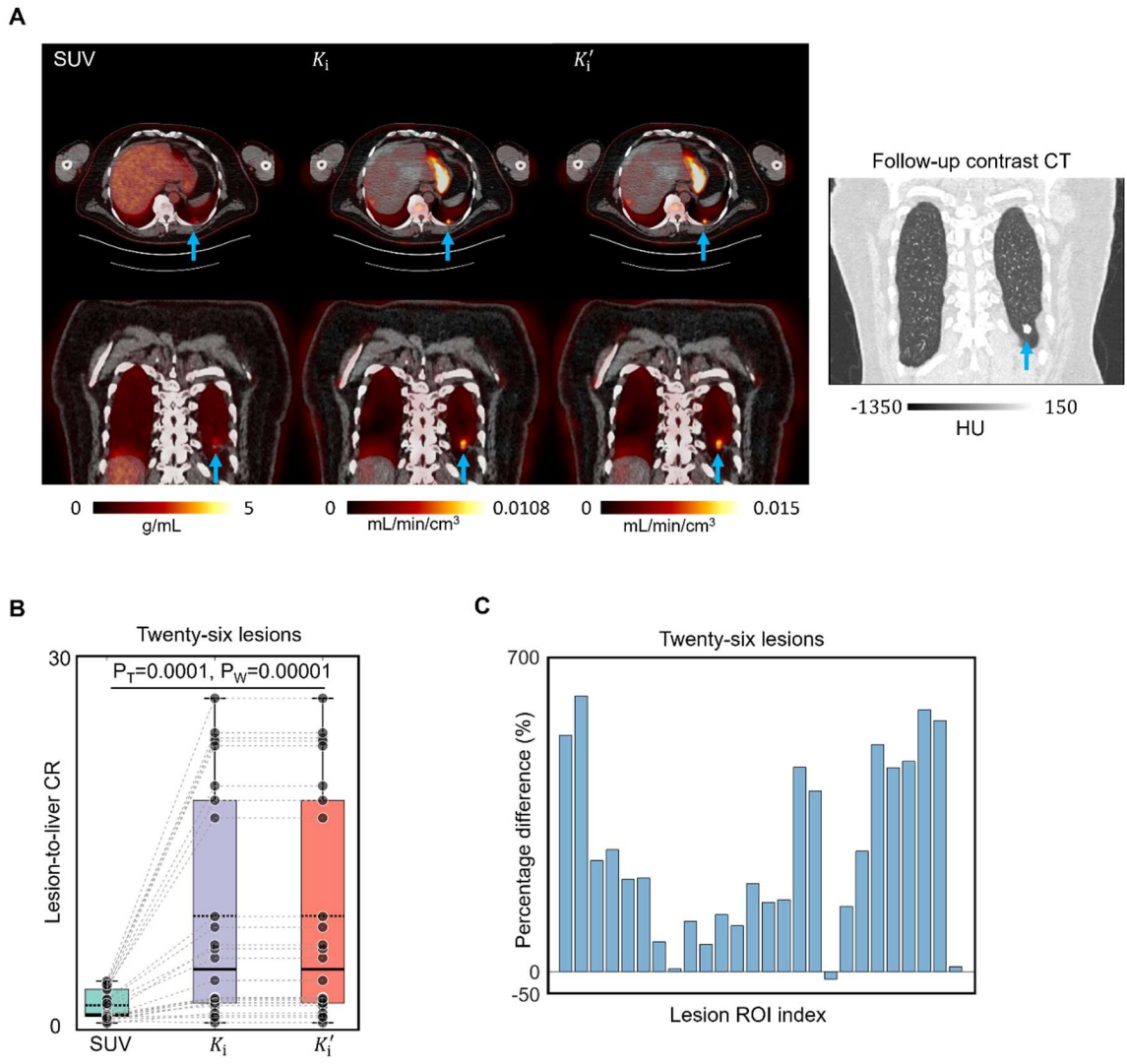

**FIGURE 4**. (A) An image comparison of lesion contrast for SUV, $K_i$ and $K_i'$. The images were superimposed on CT and shown in transverse and coronal planes. The arrow points to a lung metastasis as confirmed by a follow-up contrast CT. (B) Boxplot comparison of lesion-to-liver contrast ratio (CR) for all 26 lesions from 10 cancer patients for SUV, $K_i$ and $K_i'$. The paired lines, P values of paired t-test ($P_T$), and Wilcoxon signed-rank test ($P_W$) were included for comparing $K_i'$ and SUV. (C) The bar plot of the percent difference (%) of CR between $K_i'$ and SUV for each lesion. Note that the CR difference between $K_i'$ and $K_i$ is zero for each lesion ROI and not shown.

Figure 5A shows the SUV images of three cancer patients with a view of the heart region. Figure 5B and 5C show their corresponding standard Patlak $K_i$ and RP $K_i'$ images respectively. Even though with different intensity ranges, both $K_i$ and $K_i'$ images demonstrated a clear visualization of the myocardium (especially for the left ventricular myocardium), while the SUV images could not. This result suggests that RP parametric imaging can visualize and potentially characterize the myocardium in cancer patients.

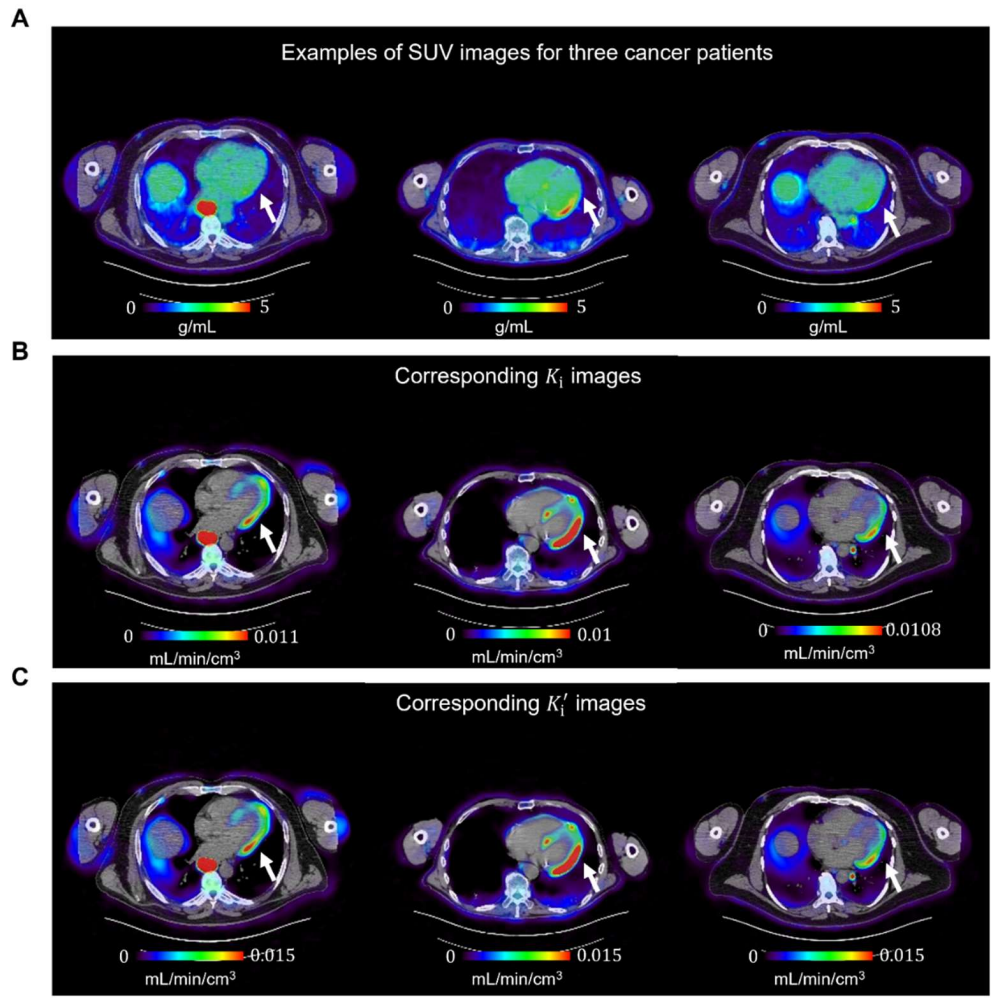

**FIGURE 5**. Comparison of parametric images with SUV images to visualize the myocardium in three cancer patients. (A) SUV, (B) standard Patlak $K_i$ and (C) RP $K_i'$ images. All images were superimposed on CT. Arrows point to the myocardium regions.

**Quantitative Potential of RP $K_i'$ across Different Subjects**

Figure 6A shows the correlation plot for 154 organ ROIs (7/subject) from all 22 subjects (10 cancer patients and 12 healthy volunteers). The inter-subject correlation between $K_i$ and $K_i'$ was strong, as indicated by a high correlation coefficient (R>0.99), a minimal P value (<0.0001), and a narrow confidence interval. The coefficient of variation for the scaling factors in these 22 subjects was 8.3% (mean of scale factors: 1.47, standard deviation: 0.12) with a 95% confidence interval of [4.8%, 11.7%].

Figure 6B shows the correlation between $K_i$ and $K_i'$ for ROI quantification of twenty-six lesions in ten cancer patients. Again, the inter-subject correlation between $K_i$ and $K_i'$ was statistically significant with a high correlation coefficient (R = 0.99) and a narrow confidence interval. The strong correlation results indicate that although not equal to $K_i$ quantitatively, the RP $K_i'$ may still enable comparable quantification across subjects.

Figure 6C shows the Bland–Altman plot of the calibrated $K_i$ from $K_i'$ as compared to the standard Patlak $K_i$. The mean difference (solid line) was close to 0 and the difference of the two measures was within the limits of agreement (dash line) for lesion ROIs. Figure 6D further shows the parametric images of the standard Patlak $K_i$ and calibrated $K_i$ for a cancer patient, as well as the difference image in absolute, demonstrating a minimal difference.

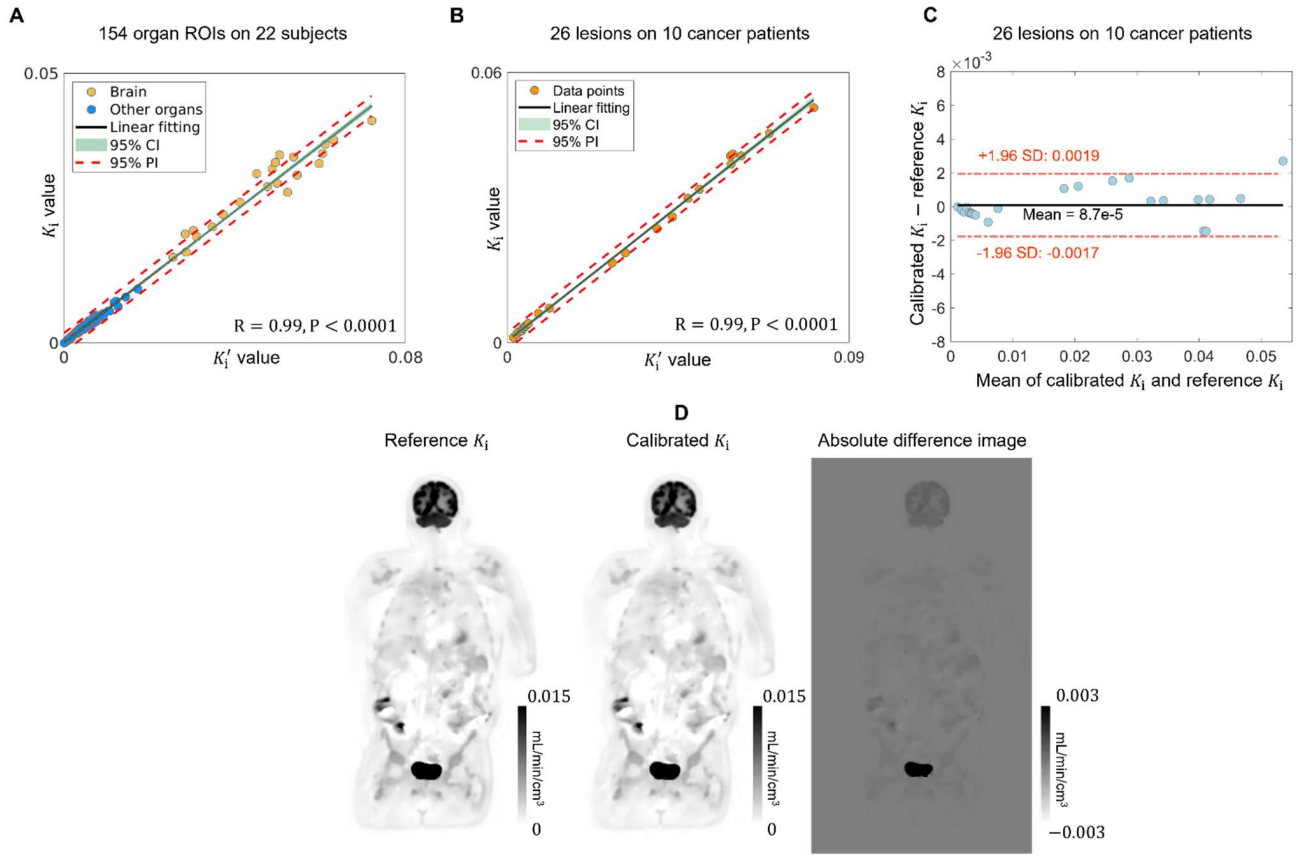

**FIGURE 6**. Demonstration of the quantitative potential of RP $K_i'$ as compared to the standard Patlak $K_i$. (A) The correlation plot between the standard $K_i$ and RP $K_i'$ for organ ROI quantification in the liver, lung, spleen, spine bone marrow, pelvic bone marrow, thigh muscle, and grey matter in 22 subjects; (B) The correlation plot between the standard $K_i$ and RP $K_i'$ for ROI quantification of 26 lesions from 10 cancer patients. The confidence interval (CI), prediction interval (PI), correlation coefficient (R), and P-value were also included. (C) The Bland–Altman plot of lesion ROI quantification for 10 cancer patients between reference $K_i$ and calibrated $K_i$ from RP $K_i'$. (D) Comparison of the parametric image of reference $K_i$ and calibrated $K_i$ of RP for a cancer patient. Their absolute difference image is included.

## Application of RP Parametric Imaging to Clinical 20-minute Scans

Figure 7 shows the total-body RP parametric images generated from two clinical 20-minute clinical FDG scans, one for a lymphoma patient scanned from 60-80 minutes and the other for a lung cancer patient scanned from 120 to 140 minutes. Their SUV images are also shown for comparison. While the RP $b'$ image exhibited comparable information as the SUV image, the $K_i'$ images showed improved lesion contrast and better visualization of the myocardium. Table 1 further shows a quantitative comparison of lesion-to-liver contrast for 13 lesions in the two patients. The lesion contrast improvement was statistically significant (P=0.001).

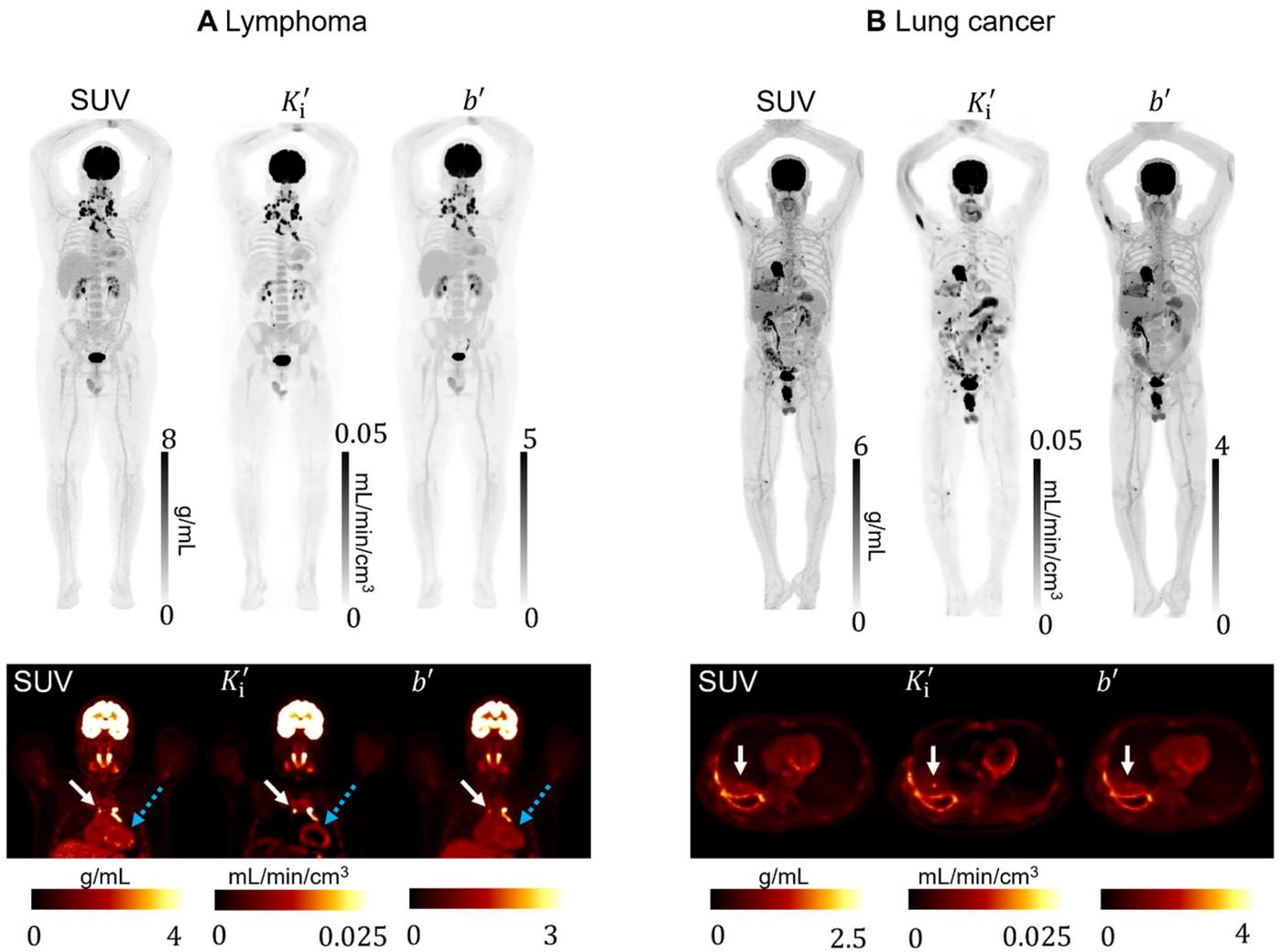

**FIGURE 7**. Total-body RP parametric imaging for two clinical scans and comparison to the SUV images. (A) A lymphoma patient scanned from 60-80 minutes post-injection. (B) A lung patient scanned from 120-140 minutes post-injection. Images of the top row are shown with maximum intensity projection. A more detailed comparison for lesions or myocardium is shown in the bottom row. The solid arrows indicated an improved lesion contrast by $K_i'$ and the dashed arrows demonstrated the ability of $K_i'$ for better myocardium visualization.

**TABLE 1** Comparison of lesion contrast for thirteen lesions in the two patients between SUV and $K_i'$.

|  | SUV | $K_i'$ | $P_T$ |
|---|---|---|---|
| Lesion contrast | 3.0 ± 1.9 | 11.0 ± 8.9 | 0.001 |

**DISCUSSION**

In this work, we demonstrated an efficient total-body parametric imaging approach using the RP plot that can be generated from a standard clinical acquisition and does not need the early-time input function. With the deep-kernel noise-reduction strategy, it becomes feasible to generate RP parametric images from a 20-minute scan (Figure 1 and Supplemental Figure 2). Compared to the earlier work (25), this paper also demonstrated that the RP intercept $b'$ is equivalent to SUVr at the time $t^*$ (Eq. 5 and Figure 3), which offers a better understanding of the theoretical aspects of the RP plot.

The results of total-body parametric images from this work further verified that the RP slope $K_i'$ image is equivalent to the standard Patlak $K_i$ image multiplied by a global scaling factor in each subject. This equivalence makes $K_i'$ equal in utility to $K_i$ for those tasks that are generally not affected by a global factor, such as lesion detection and tumor volume segmentation. Compared to SUVs, a vital benefit of $K_i'$ (and $K_i$) was the improved lesion contrast (Figure 4). Additionally, while the SUV images acquired at 1-h post-injection were unable to show the myocardium clearly in 3 out of 10 in our cancer cohort, the $K_i'$ image enabled better visualization of the myocardium for potential cardiac assessment (Figure 5), which has the potential to assess cardiotoxicity in cancer treatments (e.g., chemotherapy or immunotherapy). Addressing how to mitigate the impacts of cardiac motion is a potential challenge to explore.

Furthermore, our pilot study of 22 subjects demonstrated a strong correlation between $K_i$ and $K_i'$ across subjects for ROI quantification of lesions and major organs (Figure 6). The coefficient-of-variation of the scaling factor was relatively small (8.3%) in this cohort. These results suggest the potential to use RP $K_i'$ as a quantitative metric and warrant a future study with a large sample size to assess the impact of global scaling and evaluate the potential of $K_i'$ for therapeutic response assessment.

We further demonstrated the feasibility and benefits of applying the proposed total-body RP parametric imaging approach to standard clinical scans (Figure 7). The improved lesion contrast and myocardial visualization may facilitate the potential integration of this efficient parametric imaging approach into standard clinical workflows. Of note, the RP $K_i'$ is not aiming to replace SUVs but rather provides additional useful information without adding imaging time and scan costs.

Not limited to total-body PET, the RP parametric imaging method can also be extended to conventional shorter scanners using a multi-bed and multi-pass strategy (9,17). The noise in the parametric images may be a challenge but could be overcome by advanced reconstruction methods (35). Another application of the proposed approach would be for pediatric parametric imaging because a long (e.g., 1-hour) dynamic scan is generally impractical for pediatric patients. A short scan with a population-based input function method may be used but is challenging because of the lack of a representative dataset for this population. In contrast, the RP can be a feasible solution and will be explored in the future.

There are limitations with this work. The scan duration of dynamic data was 20 minutes with 5 minutes for each frame. There is a growing trend towards shorter clinical scans, such as 10 minutes or less (*36–38*). It would be possible to extend the current method for 10-minute scans, but further development for noise suppression and evaluation of the best possible framing protocol is required. The use of advanced noise reduction algorithms may potentially introduce texture patterns in RP parametric images. It is thus worth further assessing the reliability of the generated images, such as for quantitative lesion detectability using a physician-observer in future studies. The lesion contrast ratio was calculated using the liver as the background, not regional backgrounds that are more specific to the lesions, though the latter approach has its own limitations. In addition, the pilot study of 22 scans indicates a relatively small variation in the scaling factor $\alpha$ across subjects, thus demonstrating the quantitative potential of RP $K_i'$. However, this study was limited to a single center and did not include a test-retest component. The variation may become larger in multi-center studies in which the injection protocols could be different in different centers. A future study would be needed to explore more in this direction.

## CONCLUSIONS

In this paper, we have developed and implemented an efficient total-body parametric imaging approach using the RP plot and self-supervised deep-kernel noise reduction for dynamic FDG scans of 20 minutes duration acquired on the uEXPLORER total-body PET scanner. The RP $K_i'$ was highly correlated with standard Patlak $K_i$ for ROI quantification across subjects, demonstrating a strong quantitative potential. The method can be used to enable parametric imaging from routine clinical scans and has the potential to be applied to late-time scans and to produce parametric images for pediatric patients who cannot tolerate a long dynamic scan duration.

## DISCLOSURE

## ACKNOWLEDGMENTS

We acknowledge the contributions of team members in the EXLORER Molecular Imaging Center, UC Davis. This work was supported in part by NIH grants R01 EB033435 and R01 CA206187. The study was also supported by the In Vivo Translational Imaging Shared Resources with funds from NCI P30CA093373.

## KEY POINTS

QUESTION: Current parametric imaging with dynamic FDG-PET commonly uses a scan duration of 30-60 minutes. Is there a way to achieve total-body parametric imaging from clinical scans that acquire data for only 20 minutes starting one-two hours post tracer injection?

PERTINENT FINDINGS: The Relative Patlak plot, in combination with a deep kernel noise reduction method, was shown to be capable of generating high quality parametric images from a 20-minute clinical scan and showed benefits over the standard SUV images.

IMPLICATIONS FOR PATIENT CARE: The proposed method offers a new solution to achieve parametric imaging from a scan duration that is similar to current clinical static scans. It has the potential to apply to late-time scans and for parametric imaging of pediatric patients who cannot tolerate a long dynamic scan duration.

## REFERENCES

1. Wang G, Rahmim A, Gunn RN. PET Parametric Imaging: Past, Present, and Future. *IEEE Trans Radiat Plasma Med Sci*. 2020;4:663-675.

2. Rahmim A, Lodge MA, Karakatsanis NA, et al. Dynamic whole-body PET imaging: principles, potentials and applications. *Eur J Nucl Med Mol Imaging*. 2019;46:501-518.

3. Carson RE. Tracer kinetic modeling in PET: basic science and clinical practice. *Positrion Emission Tomography*. 2003:147-179.

4. Dimitrakopoulou-Strauss A, Pan L, Sachpekidis C. Kinetic modeling and parametric imaging with dynamic PET for oncological applications: general considerations, current clinical applications, and future perspectives. *Eur J Nucl Med Mol Imaging*. 2021;48:21-39.

5. Patlak CS, Blasberg RG. Graphical evaluation of blood-to-brain transfer constants from multiple-time uptake data. Generalizations. *Journal of Cerebral Blood Flow and Metabolism*. 1985;5:584-590.

6. Zaker N, Kotasidis F, Garibotto V, Zaidi H. Assessment of Lesion Detectability in Dynamic Whole-Body PET Imaging Using Compartmental and Patlak Parametric Mapping. *Clin Nucl Med*. 2020;45:E221-E231.

7. Fahrni G, Karakatsanis NA, Di Domenicantonio G, Garibotto V, Zaidi H. Does whole-body Patlak 18F-FDG PET imaging improve lesion detectability in clinical oncology? *Eur Radiol*. 2019;29:4812-4821.

8. Kaneko K, Nagao M, Yamamoto A, et al. Patlak Reconstruction Using Dynamic 18F-FDG PET Imaging for Evaluation of Malignant Liver Tumors. *Clin Nucl Med*. 2023;49.

9. Dias AH, Pedersen MF, Danielsen H, Munk OL, Gormsen LC. Clinical feasibility and impact of fully automated multiparametric PET imaging using direct Patlak reconstruction: evaluation of 103 dynamic whole-body 18F-FDG PET/CT scans. *Eur J Nucl Med Mol Imaging*. 2021;48:837-850.

10. Freedman NMT, Sundaram SK, Kurdziel K, et al. Comparison of SUV and Patlak slope for monitoring of cancer therapy using serial PET scans. *Eur J Nucl Med Mol Imaging*. 2003;30:46-53.

11. Cheebsumon P, Velasquez LM, Hoekstra CJ, et al. Measuring response to therapy using FDG PET: Semi-quantitative and full kinetic analysis. *Eur J Nucl Med Mol Imaging*. 2011;38:832-842.

12. De Geus-Oei LF, Van Der Heijden HFM, Visser EP, et al. Chemotherapy response evaluation with 18F-FDG PET in patients with non-small cell lung cancer. *Journal of Nuclear Medicine*. 2007;48:1592-1598.

13. Hoekstra CJ, Hoekstra OS, Stroobants SG, et al. Methods to monitor response to chemotherapy in non-small cell lung cancer with 18F-FDG PET. *Journal of Nuclear Medicine*. 2002;43:1304-1309.

14. Yin J, Wang H, Zhu G, Chen N, Khan MI, Zhao Y. Prognostic value of whole-body dynamic 18F-FDG PET/CT Patlak in diffuse large B-cell lymphoma. *Heliyon*. 2023;9:e19749.

15. Sharma A, Mohan A, Bhalla AS, et al. Role of Various Metabolic Parameters Derived from Baseline 18F-FDG PET/CT as Prognostic Markers in Non-Small Cell Lung Cancer Patients Undergoing Platinum-Based Chemotherapy. *Clin Nucl Med*. 2018;43:e8-e17.

16. Hoekstra CJ, Stroobants SC, Smit EF, et al. Prognostic relevance of response evaluation using [18F]-2-fluoro-2-deoxy-D-glucose positron emission tomography in patients with locally advanced non-small-cell lung cancer. *Journal of Clinical Oncology*. 2005;23:8362-8370.

17. Maurer A, Kotasidis F, Deibel A, Burger IA, Huellner MW. Whole-Body 18F-FDG PET/CT Patlak Parametric Imaging of Hepatic Alveolar Echinococcosis. *Clin Nucl Med*. 2023;48:1089-1090.

18. Cherry SR, Jones T, Karp JS, Qi J, Moses WW, Badawi RD. Total-body PET: Maximizing sensitivity to create new opportunities for clinical research and patient care. *Journal of Nuclear Medicine*. 2018;59:3-12.

19. Badawi RD, Shi H, Hu P, et al. First human imaging studies with the explorer total-body PET scanner. *Journal of Nuclear Medicine*. 2019;60:299-303.

20. Spencer BA, Berg E, Schmall JP, et al. Performance Evaluation of the uEXPLORER Total-Body PET/CT Scanner Based on NEMA NU 2-2018 with Additional Tests to Characterize PET Scanners with a Long Axial Field of View. *Journal of Nuclear Medicine*. 2021;62:861-870.

21. Alberts I, Hünermund JN, Prenosil G, et al. Clinical performance of long axial field of view PET/CT: a head-to-head intra-individual comparison of the Biograph Vision Quadra with the Biograph Vision PET/CT. *Eur J Nucl Med Mol Imaging*. 2021;48:2395-2404.

22. van Sluis J, van Snick JH, Brouwers AH, et al. Shortened duration whole body 18F-FDG PET Patlak imaging on the Biograph Vision Quadra PET/CT using a population-averaged input function. *EJNMMI Phys*. 2022;9.

23. Dias AH, Smith AM, Shah V, Pigg D, Gormsen LC, Munk OL. Clinical validation of a population-based input function for 20-min dynamic whole-body 18F-FDG multiparametric PET imaging. *EJNMMI Phys*. 2022;9.

24. van Sluis J, Yaqub M, Brouwers AH, Dierckx RAJO, Noordzij W, Boellaard R. Use of population input functions for reduced scan duration whole-body Patlak 18F-FDG PET imaging. *EJNMMI Phys*. 2021;8.

25. Zuo Y, Qi J, Wang G. Relative Patlak plot for dynamic PET parametric imaging without the need for early-time input function. *Phys Med Biol*. 2018;63:aad444.

26. Boellaard R, Delgado-Bolton R, Oyen WJG, et al. FDG PET/CT: EANM procedure guidelines for tumour imaging: version 2.0. *Eur J Nucl Med Mol Imaging*. 2015;42:328-354.

27. Zhang X, Xie Z, Berg E, et al. Total-body dynamic reconstruction and parametric imaging on the uexplorer. *Journal of Nuclear Medicine*. 2020;61:285-291.

28. Sari H, Mingels C, Alberts I, et al. First results on kinetic modelling and parametric imaging of dynamic 18F-FDG datasets from a long axial FOV PET scanner in oncological patients. *Eur J Nucl Med Mol Imaging*. 2022;49:1997-2009.

29. Li S, Wang G. Deep Kernel Representation for Image Reconstruction in PET. *IEEE Trans Med Imaging*. 2022;41:3029-3038.

30. Buades A, Coll B, Morel JM. A non-local algorithm for image denoising. *Proceedings - 2005 IEEE Computer Society Conference on Computer Vision and Pattern Recognition, CVPR 2005*. 2005;II:60-65.

31. Wang G, Qi J. PET image reconstruction using kernel method. *IEEE Trans Med Imaging*. 2015;34:61-71.

32. Loening AM, Gambhir SS. AMIDE: A Free Software Tool for Multimodality Medical Image Analysis. *Mol Imaging*. 2003;2:131-137.

33. Hashimoto F, Ohba H, Ote K, Kakimoto A, Tsukada H, Ouchi Y. 4D deep image prior: Dynamic PET image denoising using an unsupervised four-dimensional branch convolutional neural network. *Phys Med Biol*. 2021;66.

34. Wang G, Nardo L, Parikh M, et al. Total-Body PET Multiparametric Imaging of Cancer Using a Voxel-wise Strategy of Compartmental Modeling. *Journal of Nuclear Medicine*. 2021;7:jnumed.121.262668.

35. Hashimoto F, Onishi Y, Ote K, Tashima H, Reader AJ, Yamaya T. Deep learning-based PET image denoising and reconstruction: a review. *Radiol Phys Technol*. 2024;17:24-46.

36. Hu P, Zhang Y, Yu H, et al. Total-body 18F-FDG PET/CT scan in oncology patients: how fast could it be? *Eur J Nucl Med Mol Imaging*. 2021;48:2384-2394.

37. Ng QKT, Triumbari EKA, Omidvari N, Cherry SR, Badawi RD, Nardo L. Total-body PET/CT – First Clinical Experiences and Future Perspectives. *Semin Nucl Med*. 2022;52:330-339.

38. Hu H, Huang Y, Sun H, et al. A proper protocol for routine 18F-FDG uEXPLORER total-body PET/CT scans. *EJNMMI Phys*. 2023;10.

**Supplemental Data**

**SUPPLEMENTAL TABLE 1**
Basic information on individual subjects

| Subject | Age (year) | Sex | BMI (kg/m$^2$) | Scan protocol |
|---|---|---|---|---|
| GUC01 | 64 | M | 32 | 0-60 minutes |
| GUC02 | 61 | M | 26.3 | 0-60 minutes |
| GUC03 | 64 | M | 25.3 | 0-60 minutes |
| GUC04 | 76 | M | 20.1 | 0-60 minutes |
| GUC05 | 70 | M | 24.3 | 0-60 minutes |
| GUC06 | 73 | M | 25.6 | 0-60 minutes |
| GUC07 | 56 | M | 25.7 | 0-60 minutes |
| GUC08 | 75 | M | 21 | 0-60 minutes |
| GUC09 | 82 | F | 35.5 | 0-60 minutes |
| GUC10 | 65 | F | 18.3 | 0-60 minutes |
| HS01 | 78 | M | 24.4 | 0-60 minutes |
| HS02 | 62 | M | 29.5 | 0-60 minutes |
| HS03 | 63 | M | 24 | 0-60 minutes |
| HS04 | 73 | M | 23.8 | 0-60 minutes |
| HS05 | 60 | F | 26.4 | 0-60 minutes |
| HS06 | 67 | F | 25.5 | 0-60 minutes |
| HS07 | 67 | M | 23.3 | 0-60 minutes |
| HS08 | 61 | F | 20.2 | 0-60 minutes |
| HS09 | 60 | F | 33.7 | 0-60 minutes |
| HS10 | 61 | F | 23.1 | 0-60 minutes |
| HS11 | 64 | M | 25.2 | 0-60 minutes |
| HS12 | 58 | M | 37.6 | 0-60 minutes |
| Other01 | 42 | M | 35.4 | 60-80 minutes |
| Other02 | 71 | M | 20.5 | 120-140 minutes |

GUC: genitourinary cancer HS: Healthy subject
Other01: lymphoma patient, Other02: Lung cancer patient

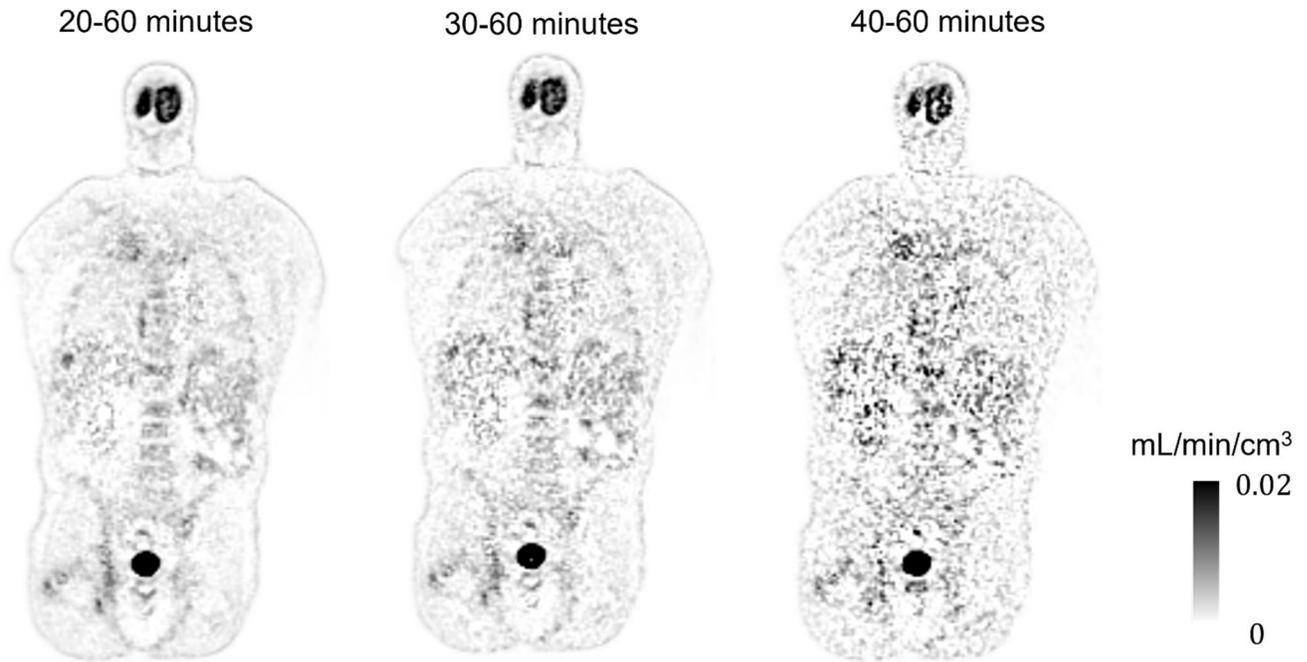

**SUPPLEMENTAL FIGURE 1.** OSEM-based $K_i'$ images generated from dynamic scans of 20-60 minutes, 30-60 minutes, and 40-60 minutes. The noise level is higher when the scan duration is reduced.

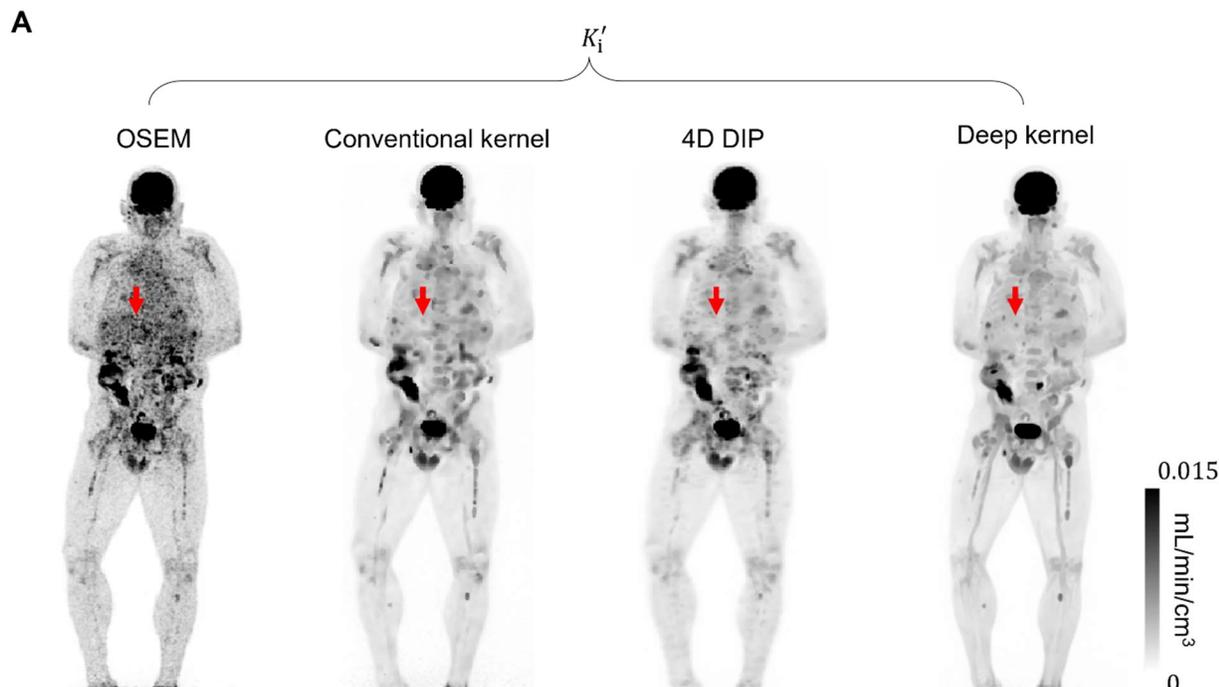
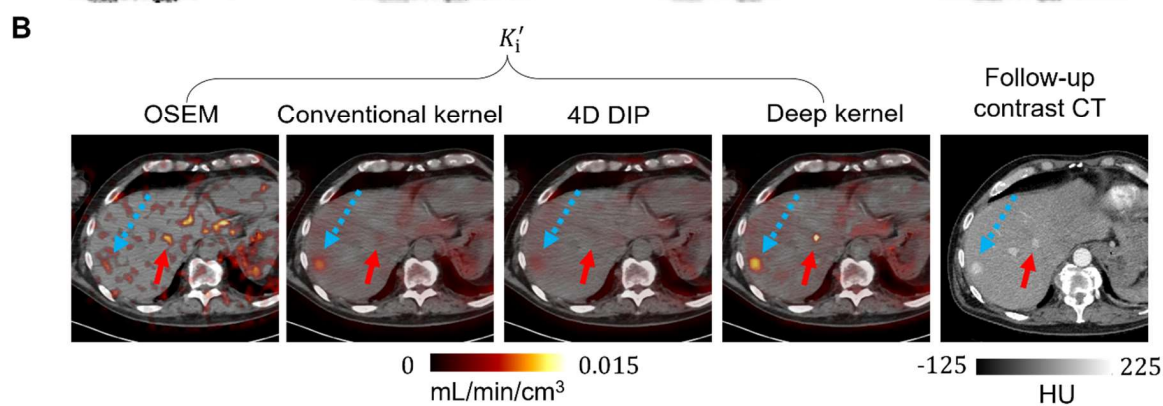
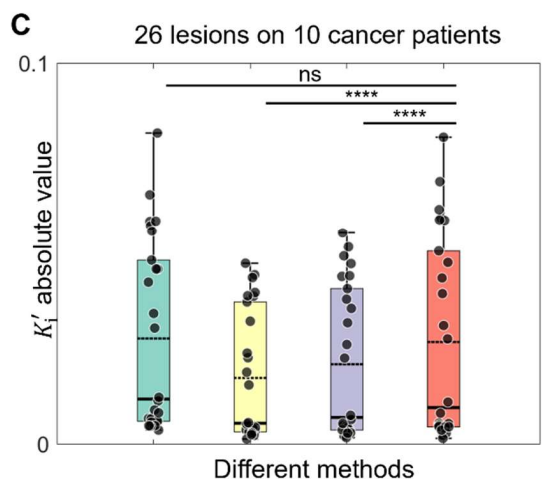
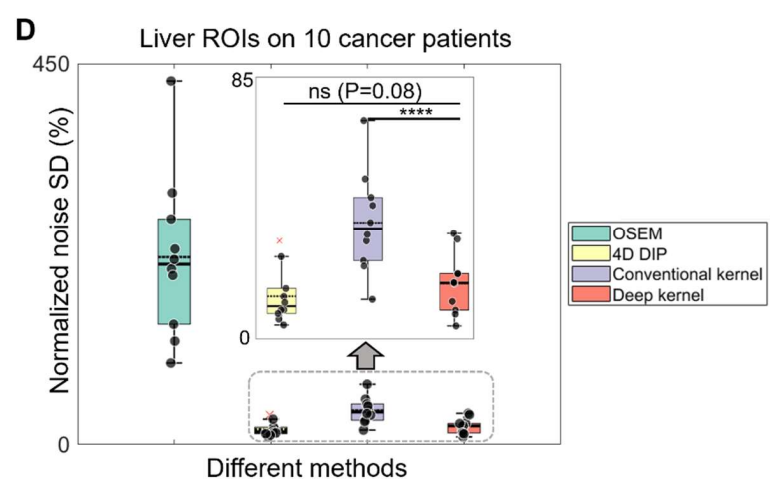

**SUPPLEMENTAL FIGURE 2**. Comparison of different methods for generating an RP $K_i'$ image. (A) Maximum intensity projection of $K_i'$ parametric images generated using four methods. A liver lesion is pointed by the red arrow. (B) Comparison for a transverse fused slice showing liver lesions which were confirmed on a follow-up contrast CT. The $K_i'$ images were superimposed on CT component of the PET/CT. Group comparisons of (C) lesion $K_i'$ values and (D) background noise standard deviation on 10 cancer patients by using paired t-test. ∗∗∗∗: P<0.0001, ns: P>0.05.

**Graphical Abstract**

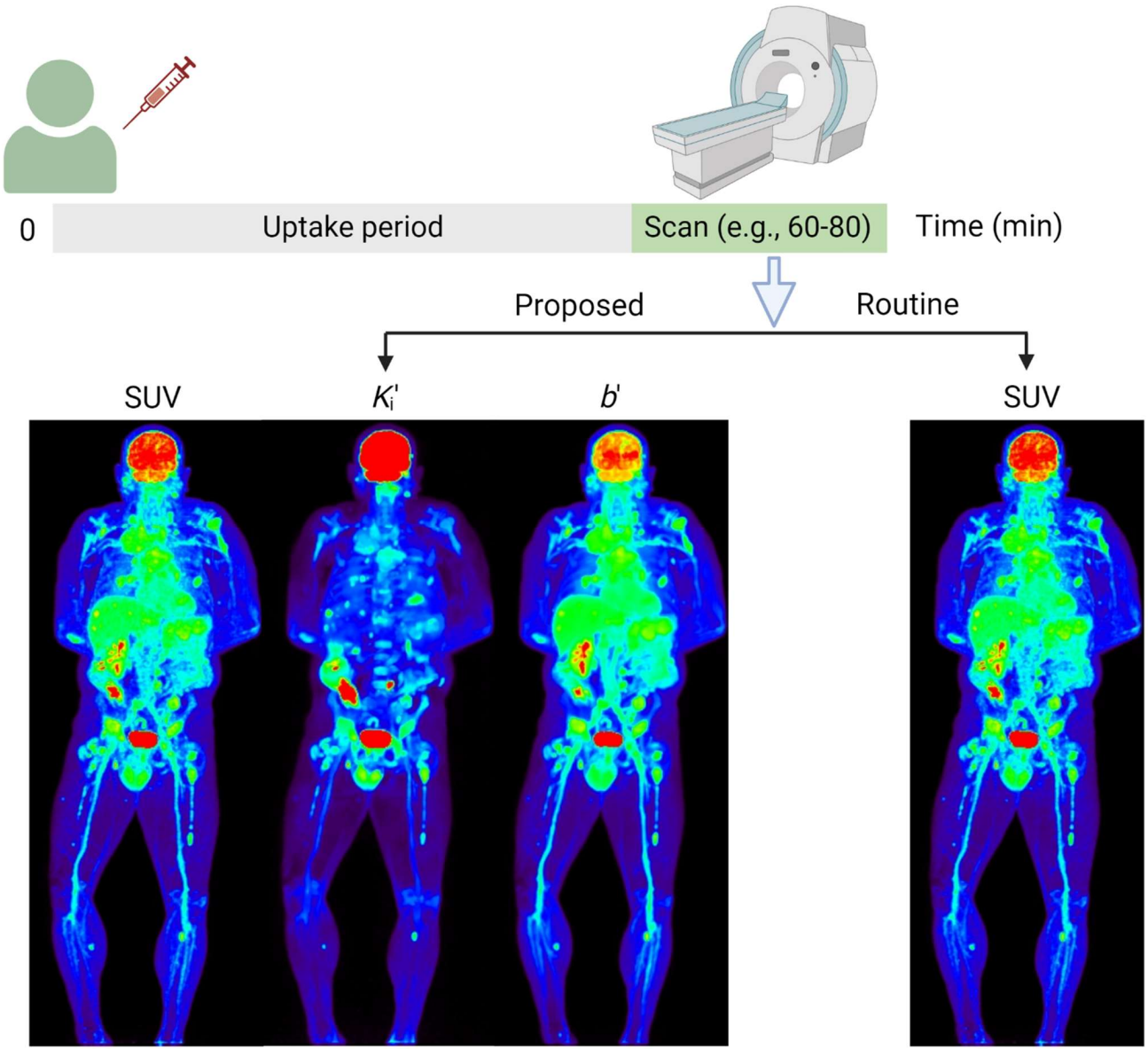